\documentclass{aa}

\newcommand{\bq}{\begin{eqnarray}} 
\newcommand{\eq}{\end{eqnarray}} 

\newcommand{\teff}{T_{\mathrm{eff}}}

\usepackage{graphicx,natbib,url}
\usepackage{float}
\usepackage{txfonts}
\begin{document}
\title{Fundamental M-dwarf parameters from high-resolution spectra
  using PHOENIX ACES models \thanks{\textbf{Based on observations carried out
    with UVES at ESO VLT.}}}

 \subtitle{I. Parameter accuracy and benchmark stars}

 \author{V. M. Passegger \inst{1}, S. Wende-von Berg \inst{1} \and A. Reiners \inst{1}}

        \institute{Institut f\"ur Astrophysik,
          Georg-August-Universit\"at G\"ottingen, Friedrich-Hund-Platz
          1, D-37077,
          Germany\\ \email{vmpasseg@astro.physik.uni-goettingen.de}
}

 \date{}


\abstract
{M-dwarf stars are the most numerous stars in the Universe; they span
  a wide range in mass and are in the focus of ongoing and planned
  exoplanet surveys. To investigate and understand their 
  physical nature, detailed spectral information and accurate
  stellar models are needed.
We use a new synthetic atmosphere model generation and
  compare model spectra to observations. To test the model accuracy, 
  we compared the models to four benchmark
  stars with atmospheric parameters for which independent information
  from interferometric radius measurements is available.
We used $\chi^2$-based methods to determine parameters from high-resolution spectroscopic observations. Our synthetic spectra are
  based on the new PHOENIX grid that uses the ACES description for the
  equation of state. This is a model generation expected to be especially
  suitable for the low-temperature atmospheres. We identified suitable 
  spectral tracers of atmospheric parameters
  and determined the uncertainties in $T_{\rm eff}$,
  $\log{g}$, and [Fe/H] resulting from degeneracies between
  parameters and from shortcomings of the model atmospheres.
  The inherent uncertainties we find are $\sigma_{T_{\rm eff}} =
  35$\,K, $\sigma_{\log{g}} = 0.14$, and $\sigma_{[Fe/H]} = 0.11$. The new model spectra achieve a reliable match to our
  observed data; our results for $T_{\rm eff}$ and $\log{g}$ are
  consistent with literature values to within 1$\sigma$. However,
  metallicities reported from earlier photometric and spectroscopic
  calibrations in some cases disagree with our results by more 
  than 3\,$\sigma$. A possible explanation are systematic errors 
  in earlier metallicity determinations that were based on insufficient 
  descriptions of the cool atmospheres. At this point, however, 
  we cannot definitely identify the reason for this discrepancy, but 
  our analysis indicates that there is a large uncertainty in the accuracy of M-dwarf 
  parameter estimates.  }
  


 \titlerunning{Fundamental M-dwarf parameters}
 \authorrunning{Passegger, Wende-von Berg and Reiners}
 \maketitle
%

\section{Introduction}

Determining atmospheric parameters in M dwarfs is very different from
the situation in Sun-like stars. The complexity of cool atmospheres is
dramatically enhanced because the main opacity sources are molecules
and not atoms. The formation and abundance of these molecules is a
complex process, and molecular absorption bands consist of thousands
of individual lines that are not at all or sometimes only poorly known. For
example, \citet{FischerValenti2005} determined effective temperature,
surface gravity, and metallicity in a sample of 1040 F-, G-, and
K-type stars. The precision they achieved in terms of $1\sigma$
uncertainties is $44$\,K for $T_{\rm eff}$, $0.06$\,dex for $\log{g}$,
and $0.03$\,dex for abundances. Accurate determinations of low-mass
star atmospheric parameters, on the other hand, still do not
provide a clear picture, in particular when metallicity is
concerned.

Previous work trying to determine stellar properties for low-mass stars 
generally dealt with each parameter separately.
\citet{RojasAyala2012} investigated near-infrared K-band spectra 
of 133 M dwarfs. They determined effective temperatures by using the 
H2O-K2 index that quantifies absorption due to H2O opacity. For calibration 
they used BT-Settl models \citep{Allard2011} of solar metallicity. The 
uncertainties in temperature lie below $100$\,K.
Another approach was used by \citet{Boyajian2012}, who calculated the 
effective temperature for nearby K and M dwarfs through interferometrically 
determined radii and bolometric fluxes from photometry. All uncertainties lie 
below 80 K. The same approach was used by \citet{vanBelleBraun2009}. They 
calculated effective temperatures for a wide range of spectral types, 
including two main-sequence M dwarfs. 
\citet{GaidosMann2014} used PHOENIX model atmospheres (\citep{Hauschildt1997, 
Allard2001}) to determine the effective temperature from spectra 
observed in the visible wavelength range. For spectra taken in near-infrared 
K band they used spectral curvature indices.
Different methods have been used to determine the surface gravity. 
\citet{Segransan2003} used interferometry to determine the angular diameter of the 
stars: together with mass-luminosity relations, the mass can be derived and the 
surface gravity can be easily calculated. Other authors, for
example, \citet{Rice2015}
and \citet{delBurgo2013}, used model fits to directly determine the surface gravity, 
which avoids assumptions about radius, mass, and age.
Determining the metallicity in M dwarfs is more difficult than 
determining temperature and surface gravity. After several decades of research, 
different methods and models can still give contradicting results, hence 
we decided to discuss this topic in more detail below.

The measurement of metallicities in M dwarfs is different to the
determination in Sun-like stars because the dense forest of spectral
features complicates a line-by-line approach. Instead, a full
spectral synthesis of the considered spectral range is preferable,
which is generally much more complex than computing
individual lines. Early attempts to measure metallicities for M dwarfs
date back to \citet{Mould1976}, who performed a line-by-line analysis
of atomic near-IR lines. \cite{Jones1996} used PHOENIX model spectra
\citep{Hauschildt1999, Allard2001} to follow a similar approach, and
\citet{Gizis1997} matched low-resolution optical spectra to the same
models. One of the first analyses of a high-resolution M dwarf
spectrum was performed by \citet{Valenti1998} and
\citet{ZborilByrne1998}, trying to fit high-resolution spectra to
PHOENIX models. They concluded that their M-dwarf metallicities were
only indicative, a conclusion that is probably valid for all previous
references. As an example for the problems inherent to M-dwarf
metallicity determinations, we can compare the values for one star
found with different analyses; for the mid-M object Gl~725B (M 3.5),
\citet{Valenti1998} reported [Fe/H]~=~$-0.92,$ while
\citet{ZborilByrne1998} found [Fe/H]~=~$-0.15$. The results for the
probably best-known mid-M dwarf, Barnard’s star (GJ~699, M4), which we
also investigated here, are spread in the literature between 
[Fe/H]~=~$-0.39\pm0.17 $ from \citet{RojasAyala2012} and $-0.75$ 
from \citet{Jones1996}. \citet{RojasAyala2012} analysed the 
equivalent widths of NaI and CaI and calibrated their scale using metallicities 
of 18 FGK+M binary systems.

An important indication that metallicity severely affects the energy
distribution in low-mass stars was provided by \cite{Delfosse2000}.
The authors used direct mass determinations from measurements in
binary systems to derive an empirical mass-luminosity relation. They
found that while this relation shows relatively little scatter in the
$K$ band, the scatter in the $V$ is huge. More specifically, the
colour index $V-K$ shows large scatter as a function of mass, which
can be explained by a scatter in metallicities, as was shown using
predictions of $V-K$ colours calculated from the PHOENIX
models. Clearly, metallicity plays a crucial role for the spectra at
visual wavelengths, while it is not as important at near-IR
wavelengths.

A quantitative improvement to metallicity calculations in M dwarfs was
achieved when \cite{Segransan2003} reported interferometric radius
measurements of M dwarfs. Direct radius measurements together with the
luminosity of and the distance to a star provide independent
constraints to effective temperature by removing one free
parameter. Furthermore, the mass-luminosity relation from
\cite{Segransan2003} can estimate the mass as a function of
the near-IR luminosity reasonably well, so that together with the radius, the surface
gravity is known as well. In systems with direct radius measurements,
measurements of the apparent luminosity, and accurate determinations
of the distance, metallicity remains the only free parameter in the
description of a stellar atmosphere. This concept was applied to two
nearby M dwarfs by \cite{WoolfWallerstein2004} and
\cite{DawsonDeRobertis2004} using \textsc{NextGen} model atmospheres
\citep{Hauschildt1999}. \cite{WoolfWallerstein2005} measured
metallicities in 15 M dwarfs for which no direct radius measurements
were available. \cite{Bonfils2005} enriched this sample with 20
systems consisting of an F-, G-, or K-dwarf primary and an M-dwarf
secondary. These binaries are believed to have formed together so that
their metal abundance is expected to be identical, which provides the
only independent test of M-star metallicity.  A similar approach was
chosen by \cite{Bean2006a, Bean2006b}, who used five binaries to
determine the metal content of the primary and from this deduced the
metal content of the M-dwarf secondaries. The model they developed was
then used to determine the metallicity of three planet-hosting M
dwarfs. \citet{Bonfils2005} also provided a colour-metallicity
relation, from which the metal abundance of M dwarfs can apparently be
derived simply from their colours. This reduces the complex problem of
atmospheric physics in M dwarfs to a two-dimensional relation between
infrared and visual luminosities and has been used in many following
studies since, for example by \cite{Casagrande2008} and for the
calibration applied in \cite{Neves2013}. However, 
\citet{RojasAyala2012} found that the results of \citet{Bonfils2005} show 
lower metallicity values for stars with solar and super-solar metallicities.
\citet{GaidosMann2014} used empirical relations between metallicity and atomic line 
strength in the H and K band calibrated with FGK+M binary systems as described 
by \citet{Mann2013}. Their results in metallicity agree excellently
well with those of \citet{RojasAyala2012}, with mean differences of 0.03 dex.
A recent work by \citet{Maldonado2015} used pseudo-equivalent widths of optical 
spectra to determine effective temperature and [Fe/H]. They showed that the calibration 
of \citet{Casagrande2008} underestimates temperatures because 
\citet{Casagrande2008} assumed M dwarfs to be blackbodies, but M dwarfs have more flux 
in the infrared than predicted for a blackbody. For metallicities, however, 
\citet{Maldonado2015} reported good agreement of their results with \citet{Neves2012} 
and \citet{Neves2014}. 

A surprising result that arose from the comparison between
metallicities in Sun-like planet-hosting stars and M-dwarf planet-hosting
stars was made by \citet{Bean2006a}, who claimed that Sun-like planet 
hosts tend to have high metal abundances, but this is not the case for M-dwarf planet hosts. This result was disproved by several subsequent 
works (e.g.\,\cite{JohnsonApps2009, RojasAyala2012, GaidosMann2014}), 
which showed that M-dwarf planet hosts also have 
solar to super-solar metallicities. 
One way to explain the discrepancy found by \citet{Bean2006a} 
is that metallicity determinations for M
dwarfs are systematically underestimating the metal abundance, implying
that the colour-metallicity relations are not generally
applicable either. This alternative was tested by \cite{JohnsonApps2009}, who
compared metallicity measurements in Sun-like stars and M dwarfs in
the overlapping region where different calibration methods are
valid. They showed that the metallicities of the high-metal M dwarfs
in their sample are underestimated by as much as 0.32~dex in [Fe/H] on
average. \cite{Neves2012} reported that the method 
of \cite{JohnsonApps2009} is a good predictor for high metallicities, but 
tends to overestimate low-metallicity M dwarfs by about 0.14~dex. 
\citet{RojasAyala2012} came to the same conclusion.
\cite{SchlaufmanLaughlin2010} inferred a photometric metallicity calibration
and also found that \cite{JohnsonApps2009} overestimated metallicities.
\citet{RojasAyala2012} reported that \citet{SchlaufmanLaughlin2010} 
underestimated values for some solar and super-solar metallicity stars, but 
for stars with sub-solar metallicity they either under- or overestimate by 
up to 0.6 dex.
A method similar to the one we describe in this work was used by 
\citet{Rajpurohit2013}. They compared low- and medium-resolution spectra 
of 152 M dwarfs to BT-Settl models and applied a $\chi^2$-method to determine 
effective temperatures. \citet{Rajpurohit2014} improved this work by 
additionally determining $\log{g}$ and metallicity and interpolating 
between the model grid points for these two parameters. This was done 
for high-resolution spectra of 21 M dwarfs. 
A comparison of available metallicity determinations was
carried out by \citet{Neves2012}.
In their recent work, \citet{Mann2015} presented their temperature 
determination using optical spectra and BT-Settl models following the 
approach of \citet{Mann2014}. For metallicity they used equivalent widths 
of atomic features in near-infrared spectra together with the empirical relation 
from \citet{Mann2013} and \citet{Mann2014}. They found only slight differences to the 
metallicities reported by \citet{RojasAyala2012}, \citet{Neves2013}, and \citet{Neves2014}.
In this paper we aim for a fresh attempt to determine the atmospheric
parameters of cool stars. We take advantage of the new PHOENIX model
grid \citep{Husser2013}, which makes use of significantly advanced
microphysics relevant to M-dwarf atmospheres. In Sect.~2 we describe
the new model, and in Sect.~3 we introduce our method to derive
atmospheric properties from comparison between synthetic spectra and
observations. Observational data from benchmark stars are described in
Sect.~4, and an analysis of these stars is carried out in Sect.~5,
before we summarise our results in Sect. 6.

\section{PHOENIX ACES model atmospheres}

For the spectral fitting we used the latest PHOENIX grid ACES
\citep[see][]{Husser2013}\footnote{http://phoenix.astro.physik.uni-goettingen.de/}. 
It makes use of a new equation of state,
which accounts especially for the formation of molecules at low
temperatures. Hence it is ideally suited for the synthetisation of
cool stars spectra. The 1D models are computed in plane-parallel
geometry and consist of $64$ layers. Convection is treated in
mixing-length geometry, and from the convective velocity a
microturbulent velocity is deduced via $v_{mic}=0.5\cdot
v_{conv}$ (see \citet{Wende2009}).  Both are used to compute the 
high-resolution spectra. An overview of the model grid parameters is shown in
Table~\ref{tab:paramspace}. In all models the assumption of local
thermal equilibrium is used.

First comparisons of these models with observations show that the
quality of the computed spectra is greatly improved in comparison to
older versions. Especially the regions of the TiO bands in the optical
are now well fitted. These regions are some of the key regions to determine effective temperatures (see below). One problem in
previous PHOENIX versions was that the $\epsilon$- and $\gamma$-TiO
bands in observations could not be reproduced with the same effective
temperature \citep{Reiners2005}. This problem is now solved in the
current version of the code (see below).

\begin{table}
  \caption{Parameter space of the spectral grid. }
  \label{tab:paramspace}
  \centering %
  \begin{tabular}{ccc}
    \hline \hline & Range & Step size \\ 
    \hline 
    $T_{\rm eff}$ [K] & 2300 -- 5900 & 100 \\ 
    $\log(g)$ & \phantom{$-$}0.0 -- $+6.0$ & 0.5 \\[1ex] 
    $[Fe/H]$ & $-4.0$ -- $-2.0$ & 1.0 \\
      & $-2.0$ -- $+1.0$ & 0.5 \\
    \hline
  \end{tabular}
\end{table}

\section{Parameter determination method}

Since the parameters, $\teff$, $\log{g}$, and metallicity in M-type
stars can be strongly degenerate, it is necessary to use spectral
regions that are sensitive to one or more stellar parameters
simultaneously to break the degeneracies. To determine the 
best set of parameters for a given star, a $\chi^2$-method is appropriate. 
The details of the method used here are described in the following.

\subsection{Spectral regions}
\label{sec:specreg}

\begin{figure}
  \includegraphics[width=0.5\textwidth,bb = 60 50 750 580]{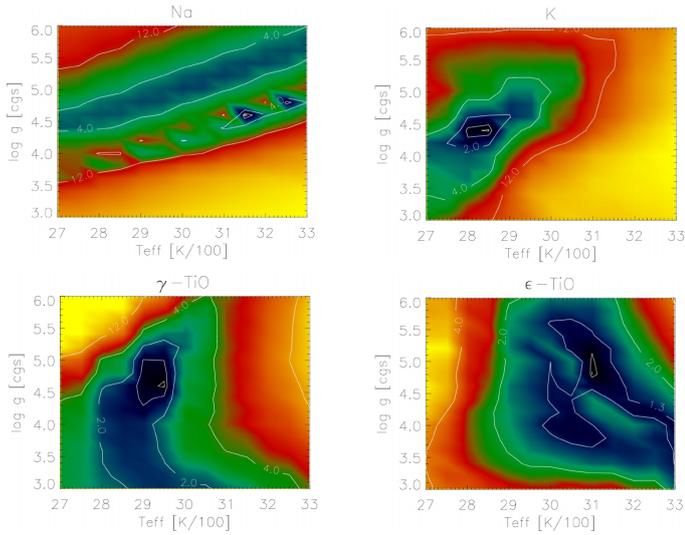}
  \caption{\label{fig:chimaplT}$\chi^2$-maps of the used oxygen bands and alkali
    lines for the M dwarf GJ551 with $\log{g}=5.02$, $T_{\rm eff}=2927$\,K, 
    $Fe/H=-0.07,$ and $S/N\sim 100$. We show the fit quality in the $T_{\rm eff}$ - $\log{g}$
    plane after calculating a rough global minimum on the low-resolution grid. }
\end{figure}

\begin{figure}
  \includegraphics[width=0.5\textwidth,bb = 60 50 750 580]{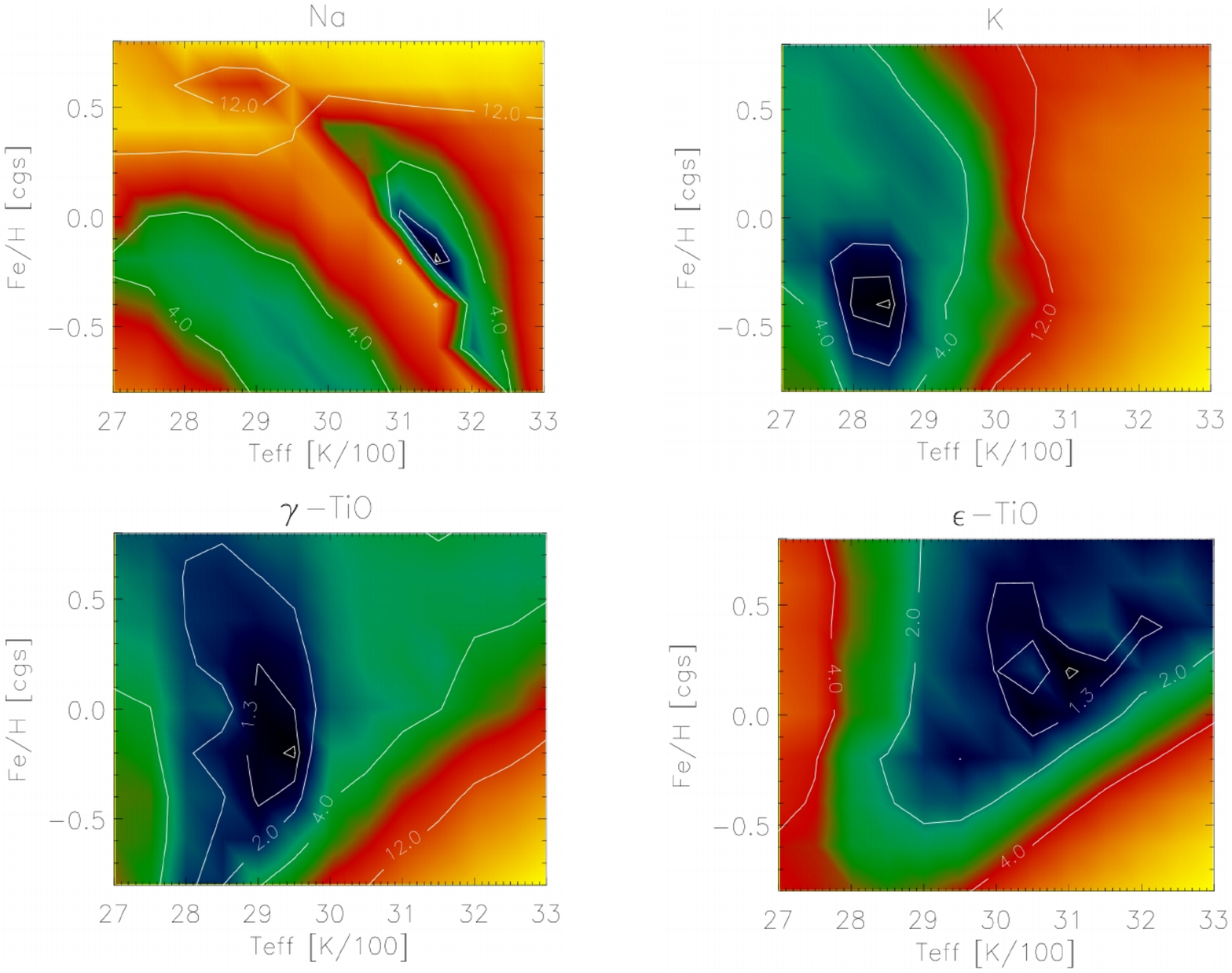}
  \caption{\label{fig:chimapmT}$\chi^2$-maps of the used oxygen bands and alkali
    lines for the M dwarf GJ551 with $\log{g}=5.02$, $T_{\rm eff}=2927$\,K,
    $Fe/H=-0.07,$ and $S/N\sim 100$. We show the fit quality in the $T_{\rm eff}$ - [Fe/H] 
    plane after calculating a rough global minimum on the low-resolution grid. }
\end{figure}

\begin{figure}
  \includegraphics[width=0.5\textwidth]{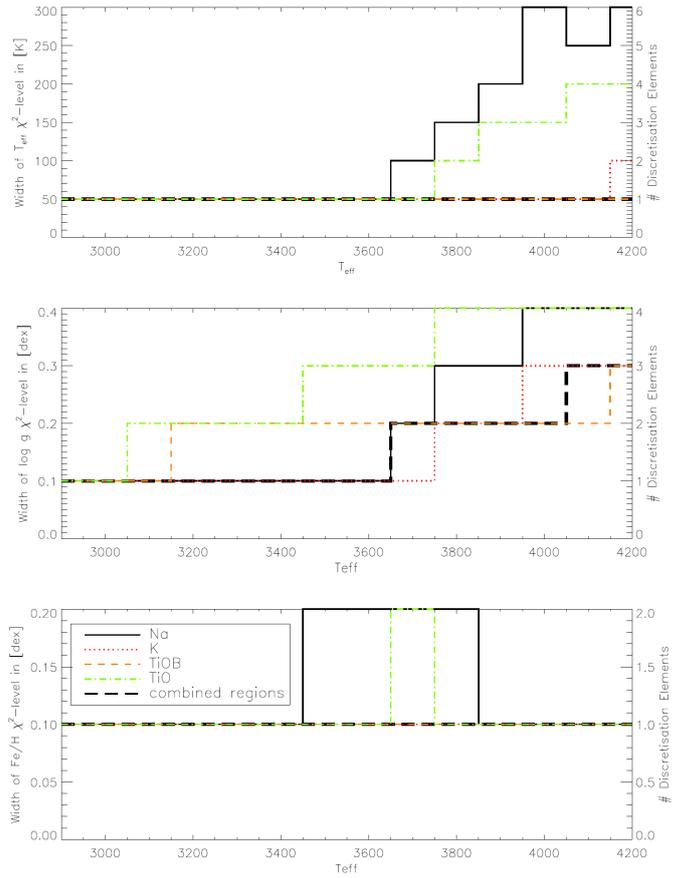}
  \caption{\label{fig:teffdependsens}Changes in sensitivity on certain 
  parameters for different lines with increasing temperature. The ciel-function is used because only discrete numbers can be resolved.}
\end{figure}

We chose several spectral regions that exhibit different dependencies
on the three stellar parameters. We chose the molecular TiO bands
around $7050$\,\AA~ and $8430$\,\AA~ ($\gamma$- and $\epsilon$-
electronic transitions, respectively) since these oxide bands are very
sensitive to effective temperature, but almost insensitive to surface
gravity (Fig.\ref{fig:chimaplT}). For cooler stars, as shown here, a
dependence on surface gravity is introduced by the increasing
micro-turbulent velocities, which enhance the line width towards lower
$\log{g}$ values (see \cite{Wende2009} for FeH as an example).  More
sensitive to surface gravity are alkali lines
(Fig.\ref{fig:chimaplT}), which show large alterations of their line
wings that are due to pressure broadening (in case of the K lines it is more
pronounced towards cooler temperatures). We chose the K- and
Na-line pairs at around $7680$\,\AA~ and $8190$\,\AA,~ respectively.
To determine the metallicity, we can use all these regions
as well because the $\epsilon$-TiO band and the alkali lines are strongly
dependent on this quantity (see Fig. \ref{fig:chimapmT}). 
\cite{Rajpurohit2014} also used the K- and Na-line pairs because of 
their sensitivity to gravity and metallicity, as well as several TiO-bands 
for temperature determination. They also illustrated the dependency on 
the different stellar parameters.
The chosen spectral regions are also affected by lines of 
the Earth's atmosphere. Especially the K- and Na-line pairs are contaminated 
by O2 and H2O bands, respectively. We used masks to exclude the 
atmospheric lines from the fit. The contamination of the TiO bands 
around $7050$\,\AA~ and $8430$\,\AA~ is very weak and can be neglected.
The degree of sensitivity on a certain parameter for a particular line
(band) also varies. In Fig.\,\ref{fig:teffdependsens} we show how the
width of a $\chi^2$-level in a $\chi^2$-map changes with varying
effective temperature (i.e. the $T_{\rm eff}$ of the synthetic
spectrum from which the $\chi^2$-map is produced is varied) for a
fixed surface gravity of $\log{g}=5.0$\,cgs and metallicity of
$z=0.0$\,cgs. The change is also measured in terms of discretion
elements in the $\chi^2$-maps, since this determines the given
resolution (left axis).  All investigated regions become less gravity sensitive towards higher $T_{\rm eff}$ , which is probably due to the
increasing thermal broadening, which becomes stronger than the pressure
broadening (see also Sect.\ref{sec:accuracy}). At this point we have to
keep in mind that the sensitivity curves only represent the
behaviour for the chosen parameter values. They look different for
other metallicities or surface gravities, which are not
shown in detail here. Nevertheless, it becomes clear that the combined
$\chi^2$-map of all regions is the best choice for most temperatures
and provides the highest sensitivity. Even though the
sensitivity of the K lines is better for most temperatures than the
combined one, we would not suggest to use only these lines because the
results would then only depend on two lines. Experience in fitting
real observations showed that the results deduced from the
K lines alone can differ from the combined solution where the latter matches
the actual parameters to a higher accuracy.

\subsection{$\chi^2$-method}

To determine the stellar parameters, we basically used a
$\chi^2$ reduction. We started with the $\chi^2$ determination using a low-resolution grid of atmospheres in a wide range around the expected
parameters of the object of interest. In this step, the models were 
first convolved to match the observed spectral resolution. Then we normalised 
the average flux of the models and the observed spectra to unity. 
After this, the models were interpolated 
to the wavelength grid of the observed spectra. Finally, every model 
of the grid was compared to the observed spectrum at each wavelength point,
and the $\chi^2$ was calculated to determine a rough global minimum. To obtain 
more pronounced minima in the $\chi^2$-map, a dynamical mask was applied to 
every model spectrum. Figures \ref{fig:chimaplT} and \ref{fig:chimapmT} show the 
$\chi^2$-maps after this first step for the different wavelength ranges we used. 
In the next step we used the 
IDL \emph{curvefit}-function to determine accurate stellar parameters 
around a narrow range of the minimum using linear interpolation 
inside the model grid.
To do this, we used the three smallest local minima from
the combined $\chi^2$-map as start values for the fitting algorithm to search
for a global minimum that may be located between the grid points of
the $\chi^2$-map.

\subsubsection{\emph{Dynamical mask} and re-normalisation}

As already mentioned, not all parts of the spectrum show the same
dependence on the stellar parameter. In certain parameter regions,
certain lines react stronger or weaker to variations. To
weight the very sensitive lines more in the $\chi^2$-determination, we
applied a dynamical mask that depends on the model spectrum  
currently applied to the observed spectrum. In a first step, we produced a
synthetic mean spectrum, constructed from all model spectra 
within the parameter range used for $\chi^2$ calculation. Then the difference between the mean spectrum and the 
model spectrum, which should be applied to the observations, was 
determined by dividing the model by the mean spectrum. Those regions showing a 
wide difference (i.e. ratios very different from 1) indicate wavelength 
regions that are very sensitive to the actual parameter configuration. 
These regions were masked and used as weighting in the $\chi^2$ calculation. 
Applying such a mask has no basic influence on finding the best fit, 
but results in more pronounced minima in the $\chi^2$-maps. 
To account for slight differences of
the continuum level and possible linear trends between the already
normalised observed and computed spectra, we applied a re-normalisation.
We used
\begin{eqnarray}
F^{obs}_{re-norm}=F^{obs}\cdot\frac{{\rm continuum~fit_{model}}}{{\rm
    continuum~ fit_{observation}}},
\end{eqnarray} 
where the continuum fits are linear. This has no significant
influence on the positions of the minima, but improves the overall
accuracy.

\subsubsection{Fit algorithm}

The IDL \emph{curvefit}-function turned out to be the most stable,
accurate, and fastest algorithm to determine the set of
parameters. We also tested the IDL \emph{powell}- and
\emph{amoeba}-algorithms. Since the \emph{curvefit}-function requires
continuous parameters, we performed a three-dimensional linear
interpolation in the computed spectra on the grid shown in
Table~\ref{tab:paramspace}. The interpolation was made for each
wavelength point in the required wavelength range.
 
\begin{figure}
\includegraphics[width=0.45\textwidth,bb = 30 10 530
  780]{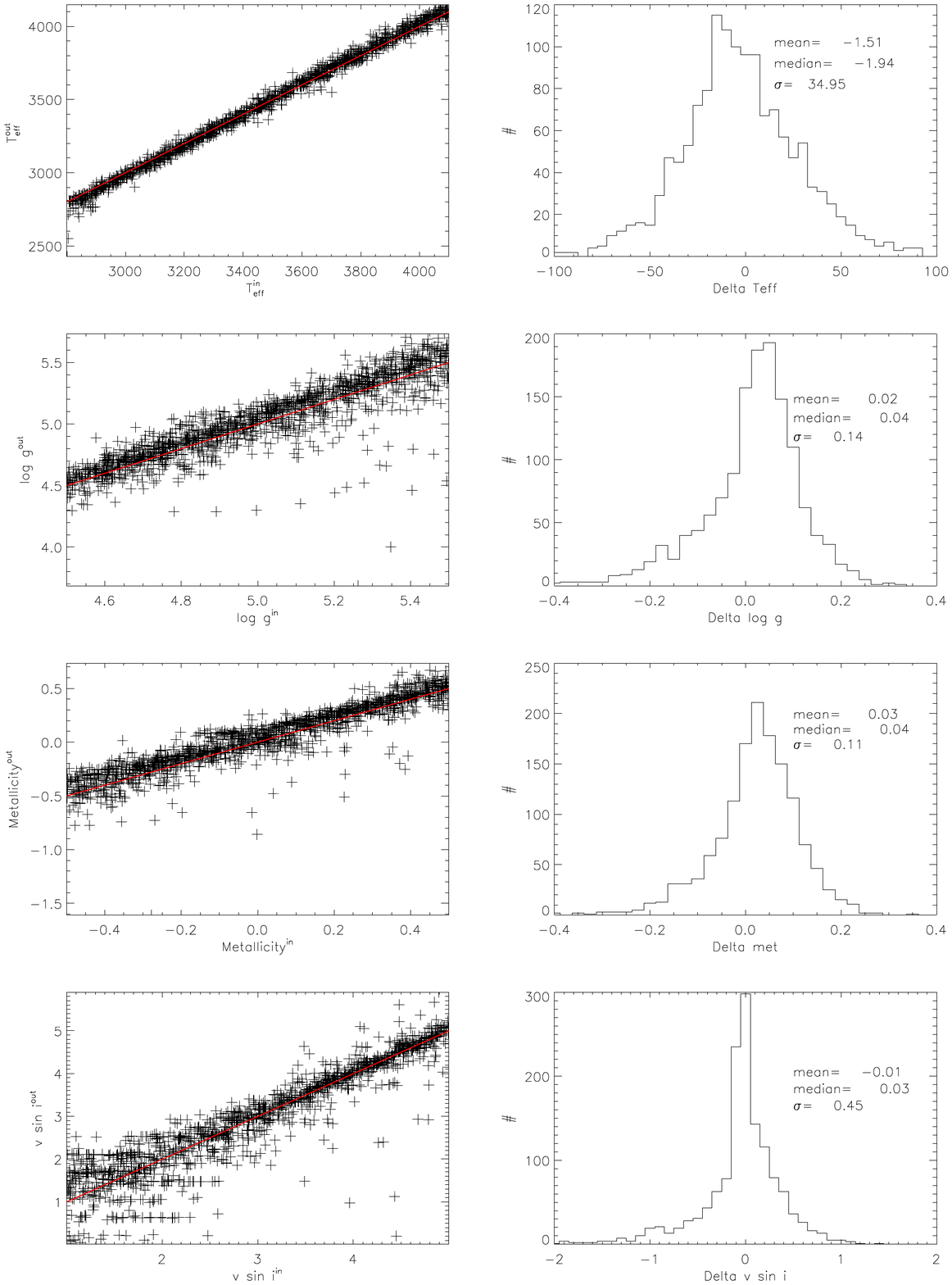}
\caption{Comparison and deviations between output and input parameters.
  Left columns: Input (not starting) parameters against
  determined parameters. The diagonal line represents a one-to-one
  correspondence. Right columns: Histograms of the deviations between
  output and input parameters.  }
\label{fig:stat_eval_I}
\end{figure}

\begin{figure*}
\includegraphics[width=0.95\textwidth,bb = 30 30 700
  580]{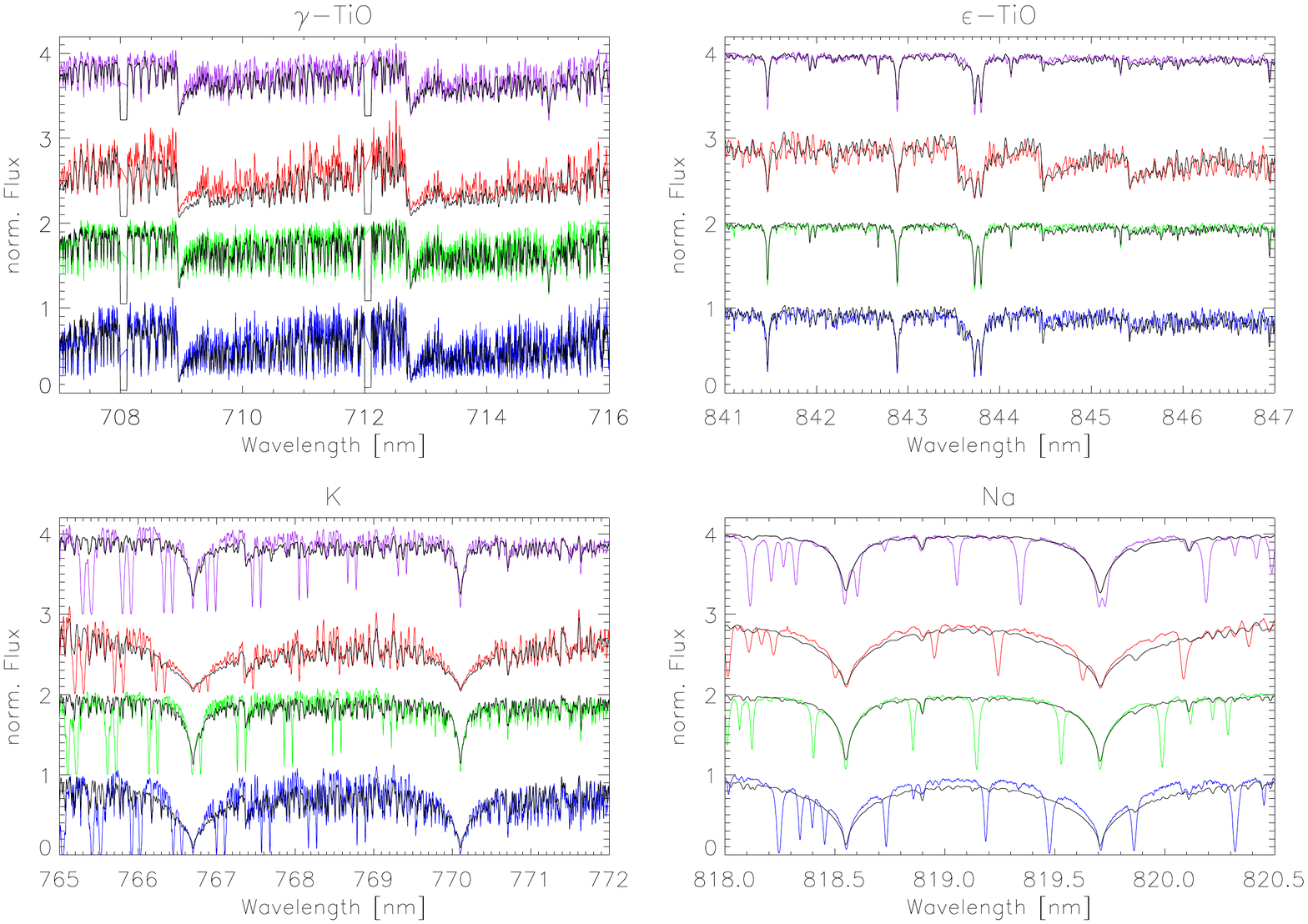}
  \caption{Comparison of the new PHOENIX models (grey) to four stars
    with known temperature and surface gravity (colour; from top to
    bottom: GJ411, GJ551, GJ667C, and GJ699).  The new models provide a
    substantial improvement over the old model generation
    \citep[see][]{Reiners2005}. In particular, the two TiO bands yield
    consistent results, although there might still be some problems 
    fitting the K- and Na-lines (e.g. see spectrum of GJ551 and GJ699). The data were not corrected for
    absorption lines from Earth’s atmosphere, therefore many sharp
    lines seen in the observations are not captured by the
    models. These lines were not included in the fit.}
\label{fig:stdM_comp}
\end{figure*}
 
\subsection{Precision of the method}
\label{sec:accuracy}

To test the precision of our method, we produced a set of
$1400$ spectra with uniformly distributed random parameters ($\teff$,
$\log{g}$, [Fe/H], and $v\sin{i}$) and different resolutions 
of $R=100\,000$, $44\,000,$ and $10\,000$. 
We added Poisson noise to simulate
a $S/N$ of $\sim 100$. The starting values differ from the actual
parameters in a range of $\pm 50$\,K for $T_{\rm eff}$, $\pm
0.25$\,dex for $\log{g}$ and [Fe/H], and $\pm 1$\,kms$^{-1}$ for
rotational broadening. We also used a continuum normalisation different to the
one used in the fitting routine to simulate
normalisation effects. We used the method described above to determine
the parameters and show the deviations from input to output parameters
in the set of histograms in Fig.~\ref{fig:stat_eval_I} for $R=100\,000$.

The deviation from a mean residual value of zero probably stems from
the fitting algorithm, since we observe a change in sign between the
\emph{curvefit}-function and the \emph{amoeba}-algorithm.  The two
algorithms use different convergence criteria, which we believe
results in the offset. In the presented case all values except
for the resolving power (or $v\,\sin{i}$) are positive, which means that the
determined parameters are slightly higher than the actual ones.

We find standard deviations from the mean value of $35$\,K for $T_{\rm
  eff}$, $0.14$\,dex for $\log{g}$, $0.11$\,dex for [Fe/H], and
$0.5$\,kms$^{-1}$ for $v\sin{i}$. No Gaussian FWHM was used since the
residual distributions are not symmetric. The reason for the asymmetry
in the $\log{g}$ and [Fe/H] plot is probably due to the logarithmic
character of the quantity. For low effective
temperatures the derived temperatures are systematically lower. This
changes towards high temperatures into the opposite. This is likely
related to the temperature-sensitive TiO bands, which are hardly
present for high temperatures and start to saturate towards the cooler
end. However, the average deviations at both ends are about $50$\,K
and agree with the final precision.

In the plot for the rotational velocities, the largest scatter is
found for values below $3$\,km. This is explained by the instrumental
resolution that we set to $R=100\,000$, which equals a broadening of
$\sim 3$\,km. Hence lower velocities cannot be expected to be
resolved.

For lower resolutions the standard deviations will increase because 
the spectra contain less information and close lines begin to overlap. 
For resolutions around $R=44\,000$ (like our observed 
spectra from GJ441 and GJ551, see next section) the 
deviations are $40$\,K for $T_{\rm eff}$, $0.18$\,dex for $\log{g}$, 
$0.14$\,dex for [Fe/H], and $1$\,kms$^{-1}$ for $v\sin{i}$. For $R=10\,000$ 
the deviations are $124$\,K for $T_{\rm eff}$, $0.46$\,dex for $\log{g}$, 
$0.36$\,dex for [Fe/H], and $7.5$\,kms$^{-1}$ for $v\sin{i}$. However, because of the resolution we do not expect to resolve rotational velocities lower 
than $\sim 7$\,kms$^{-1}$ for $R=44\,000$ and $\sim 30$\,kms$^{-1}$ for $R=10\,000$.

\section{Observational data}

\subsection{Observed data} 

The observational data used to determine the spectral
parameters are UVES spectra. The spectrum of GJ411 was taken under 
the program ID 74.B-0639 with a resolution of $R \sim 46\,000$. GJ551 was 
observed under the program ID 73.C-0138 with a resolving power of 
$R \sim 42\,300$. The spectra of GJ667C and GJ699 were taken under the 
program ID 87.D-0069. The high-resolution mode with a slit width of $0.3$''
was used, leading to a resolving power of $R \sim 100\,000$. The
observations cover a wavelength range from $640$\,nm to $1020$\,nm on
the two red chips. For GJ411 and GJ551 the ESO PHASE3 data products 
were used. The data for GJ667C and GJ699 were reduced using the 
ESOREX pipeline for UVES. The wavelength solution is based on the Th-Ar calibration
frames. All orders were corrected for the blaze function and also
normalised to unity continuum level. Then all orders were merged. For
overlapping orders (at shorter wavelengths) the red sides were used
because their quality is better. Some of our spectra contain orders merged 
into a one-dimensional spectrum; this merging of orders can potentially lead to 
artefacts caused by incorrect order merging, such as line deformation or alteration 
of line depths, which is particularly problematic if spectral features used to determine the 
parameter are influenced. Our experience is that the ESO pipelines 
perform order merging very carefully with little systematic influence. Moreover, 
we used a number of different features in our fitting procedure that are unlikely 
to systematically fall into regions close to order edges. We therefore believe 
that potential defects from individual lines and order merging are negligible. 
In a last step we removed bad pixels and cosmics.

\subsection{Objects}
\label{sect:objects}

We wish to test the results determined from our PHOENIX models 
in stars with well-studied parameters. We used four objects for
these benchmark tests. The first three are stars for which
interferometric radius determinations \citep{Segransan2003} allow an
independent estimate of their surface gravity and temperature together
with information on luminosity and distance
(Table\,\ref{tab:results}). Our fourth benchmark object, GJ~667C, was
used for a comparison of metallicity that is known for its binary
companions. This fourth object mainly serves as an additional
consistency check, a more systematic investigation considering a
larger sample of binary components will be carried out in a subsequent
paper when a larger sample of M-dwarf spectra is analysed. All four
benchmark stars are very slow rotators with no indication for
rotational line broadening. The first three objects also have measured
rotation periods consistent with very slow rotation and low activity
\cite[see][]{Reiners2012}.

\begin{itemize}

\item \textbf{GJ411 (M2):} This is the hottest star we
  investigated. With its slow rotation and no signs of high magnetic
  activity, it is an ideal target for testing the new model
  atmospheres at the warm end of M-dwarf temperatures. Radius
  measurements and a determination of $\log{g}$ are taken from
  \cite{Segransan2003}. Temperature estimates derived from
    interferometric radius and luminosity are $T_{\rm
      eff}$~=~$3593\pm$60~K \citep{vanBelleBraun2009} and $T_{\rm
      eff}$~=~$3460\pm$37~K \citep{Boyajian2012}. Results from
    spectroscopy yielded temperature estimates of $T_{\rm
      eff}$~=~$3526\pm$18~K \citep{RojasAyala2012}, $T_{\rm
      eff}$~=~$3697\pm$110~K \citep{GaidosMann2014} and 
      $T_{\rm eff}$~=~$3563\pm$60~K \citep{Mann2015}. The metallicity of
    this star was reported as [Fe/H]~=~$-0.4$
    \citep{WoolfWallerstein2005}, [Fe/H]~= ~$-0.41\pm$0.17
    \citep{RojasAyala2012}, [Fe/H]~=~$-0.35\pm$0.08
    \citep{Neves2013}, [Fe/H]~=~$-0.3\pm$0.08
    \citep{GaidosMann2014}, and most recently [Fe/H]~=~$-0.38\pm$0.08 by 
    \citet{Mann2015}.

\item \textbf{GJ699 (Barnard's star, M4):} Barnard's star is of the
  most frequently observed and well-studied stars. It is classified as an M4
  dwarf and shows no indications for enhanced magnetic activity in the
  spectra. Radii measurements are reported in
    \cite{Segransan2003} together with $\log{g}$. \cite{Boyajian2012}
    derived a temperature of $T_{\rm eff}$~=~$3230\pm$10~K from radius
    and luminosity data. Spectroscopic temperatures reported are
    $T_{\rm eff}$~=~$3266\pm$29~K \citep{RojasAyala2012}, $T_{\rm
      eff}$~=~$3338\pm$110~K \citep{Neves2014}, $T_{\rm
      eff}$~=~$3247\pm$61~K \citep{GaidosMann2014}, and 
      $T_{\rm eff}$~=~$3228\pm$60~K \citep{Mann2015}. Metallicities
    reported for this star are spread between [Fe/H]~=$-0.39$ and $-0.75$,
    recent examples are [Fe/H]~=~$-0.39\pm$0.17
    \citep{RojasAyala2012}, [Fe/H]~=~$-0.52\pm$0.08 \citep{Neves2013},
    [Fe/H]~=~$-0.51\pm$0.09 \citep{Neves2014}, 
    [Fe/H]~=~$-0.32\pm$0.08 \citep{GaidosMann2014}, and 
    [Fe/H]~=~$-0.40\pm$0.08 \citep{Mann2015}.

\item \textbf{GJ551 (M6):} This M6 star is the coolest object in our
  sample. Like the other stars, no indications of enhanced activity
  were found and the spectra are consistent with slow
  rotation. $T_{\rm eff}$ and $\log{g}$ from \cite{Segransan2003} are
  shown in Table\,\ref{tab:results}. \cite{Neves2013} derived a temperature 
  of $T_{\rm eff}$~=~$2659$\,K, which 
  is significantly lower than the value from interferometry. 
    Reports of metallicity include [Fe/H]~=~0.19 \citep{Edvardsson1993},
  [Fe/H]~=~$0.0\pm$0.08 \citep{Neves2013}, 
  [Fe/H]~=~$0.16\pm$0.20 \citep{Neves2014}, and [Fe/H]~=~$-0.03\pm$0.09
  \citep{Maldonado2015}.

\item \textbf{GJ667C (M1.5):} This M1.5V dwarf also is a slow rotator without
  signs of strong magnetic activity; it shows no H$\alpha$
  emission, for example. The star hosts several low-mass planets. Atmospheric
  parameters were reported in \cite{AngladaEscude2013}. The star has
  no interferometric radius constraints, hence we lack independent
  information on temperature and gravity. Because it is part of a
  multiple star system, we have constraints on metallicity assuming
  that the three components of the triple share similar
  metallicities. For the K3V component of the system, we found five
  independent metallicity determinations that we collect in
  Table\,\ref{tab:GJ667FeH}. The mean and median values of the sample
  are [Fe/H]~=~$-0.55$ with a standard deviation of $\sigma{\rm
    [Fe/H]} = 0.08$. For the temperature, literature data are
    $T_{\rm eff}$~=~$3350\pm$50~K \citep{AngladaEscude2013}, $T_{\rm
      eff}$~=~$3351$ \citep{Neves2013}, $T_{\rm
      eff}$~=~$3445\pm$110~K \citep{Neves2014}, and $T_{\rm
      eff}$~=~$3472\pm$61~K \citep{GaidosMann2014}. Spectroscopic
    determination of metallicity for the C component are
    [Fe/H]~=~$-0.55\pm$0.1 \citep{AngladaEscude2013},
    [Fe/H]~=~$-0.53\pm$0.08 \citep{Neves2013}, [Fe/H]~=~$-0.50\pm$0.09
    \citep{Neves2014}. \cite{GaidosMann2014} reported a value of
    [Fe/H]~=~$-0.3\pm$0.08 from an empirical relation between
    metallicity and atomic line strength, described in
    \cite{Mann2013}.
\end{itemize}

\begin{table}
  \centering
  \caption{\label{tab:GJ667FeH}Metallicities measured for GJ~667AB}
  \begin{tabular}{ll}
    \hline \hline Reference & [Fe/H] \\ \hline \cite{Perrin1988} &
    $-0.59$\\ \cite{Marksakov1988} & $-0.52$\\ \cite{Thevenin1998} &
    $-0.55$\\ \cite{Santos2005} & $-0.43$\\ \cite{Taylor2005} &
    $-0.66$\\ \hline
  \end{tabular}
\end{table}

\section{Results}

\begin{center}
  \begin{table*}
    \caption{\label{tab:results}Parameters for benchmark stars with
      interferometric radii. Columns 3--5: literature values; Cols. 6--9:
      parameters determined from our model fit.}
    \begin{tabular}{llllllll}
      \hline \hline Object& Sp& $T_{\rm eff}$ [K]$^{b}$& $\log{g}^{b}$ &
      [Fe/H]$^{a}$ & $T_{\rm eff}^{fit}$ & $\log{g}^{fit}$ &
      [Fe/H]$^{fit}$\\ \hline GJ411& M2.0& $3570 \pm 42$ &
      $4.85 \pm 0.03$ & $-0.35 \pm 0.08$ & $3565 \pm 40$ &
      $4.84 \pm 0.18$ & $ +0.00 \pm 0.14$\\ GJ699 &M4.0& $3163
      \pm 65$ & $5.05 \pm 0.09$ & $-0.52 \pm 0.08$ &
      $3218 \pm 35$ & $5.25 \pm 0.14$ & $- 0.13 \pm 0.11$\\ GJ551
      &M5.5& $3042 \pm 117$ & $5.20 \pm 0.23$ & $
      + 0.00 \pm 0.08$ & $2927 \pm 40$ & $5.02 \pm 0.18$ & $- 0.07
      \pm 0.14$\\ \hline
      $^{a}${\cite{Neves2013}}&&&&&&&\\ $^{b}${\cite{Segransan2003}}&&&&&&&\\
    \end{tabular}
  \end{table*}
\end{center}

\begin{figure}
  \includegraphics[width=0.45\textwidth,bb = 50 50 690 550]{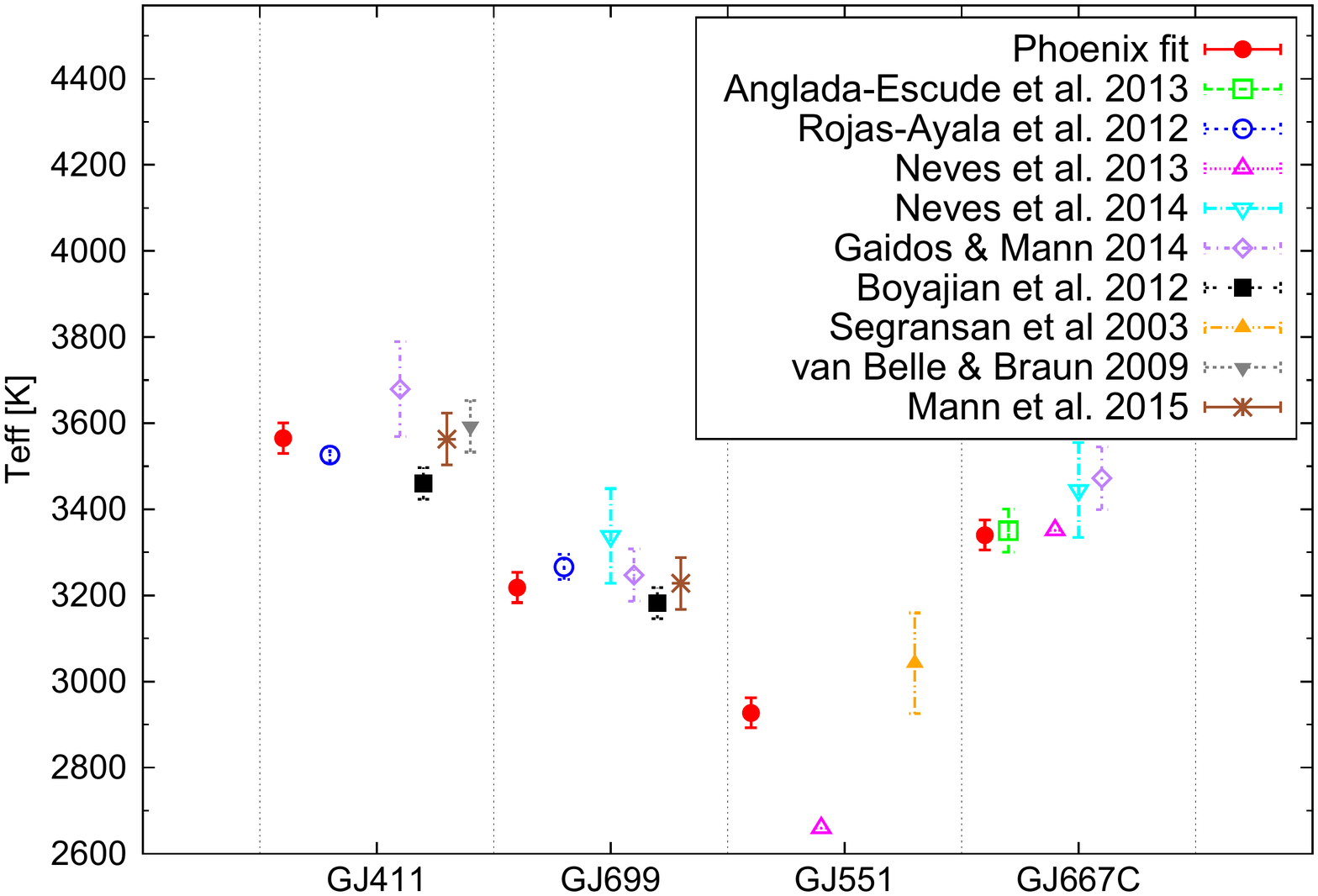}
  \includegraphics[width=0.45\textwidth,bb = 50 50 690 550]{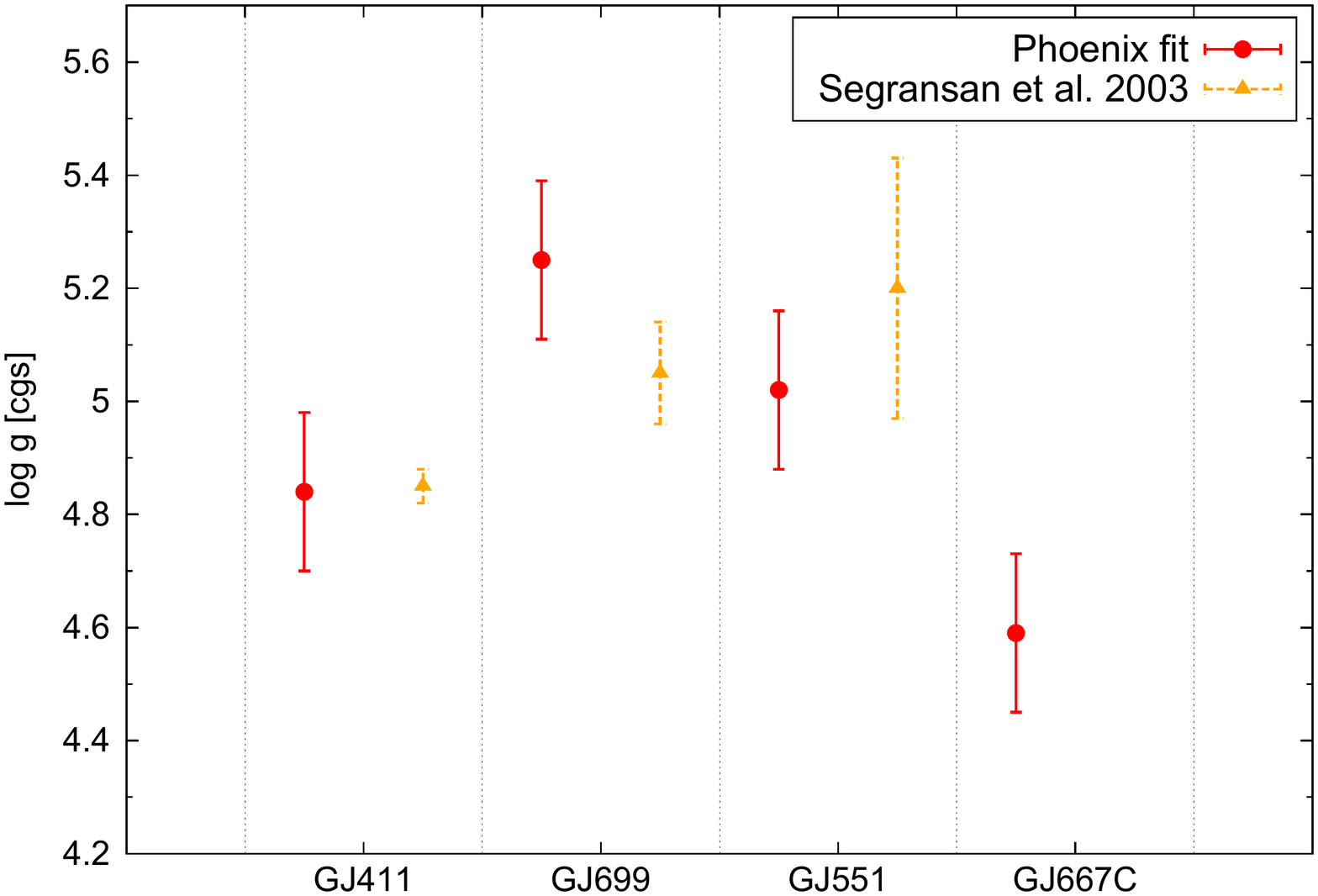}
  \includegraphics[width=0.45\textwidth,bb = 50 50 690 550]{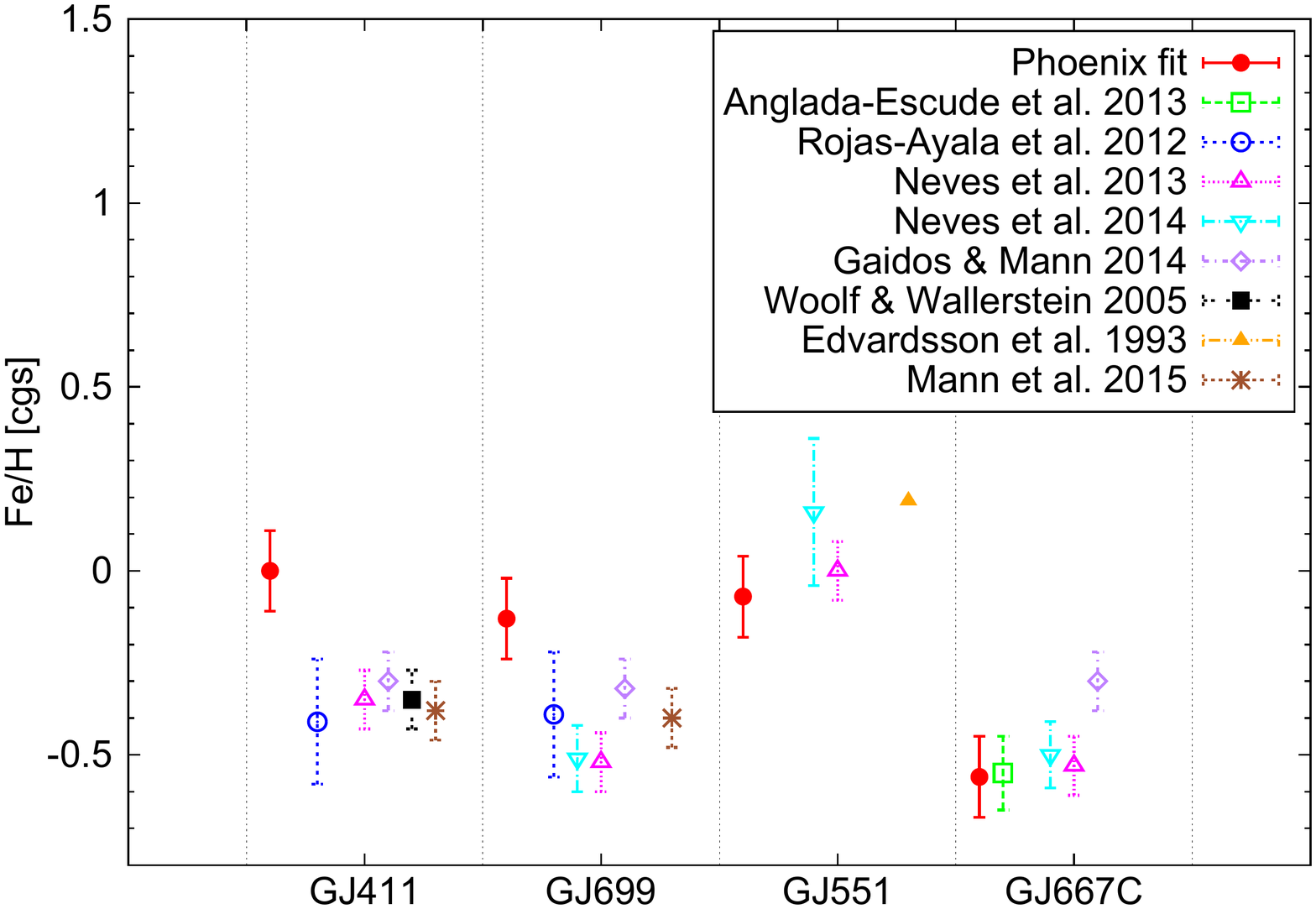}
  \caption{\label{fig:parcompare}Comparison between our Phoenix results and results reported
    in the literature values. \emph{Top panel: }Effective temperature,
    \emph{middle panel:} Surface gravity, \emph{bottom panel:}
    Metallicity.}
\end{figure}

Figure \ref{fig:stdM_comp} shows the spectra of all four stars together 
with the best-fit models.
The atmospheric parameters derived from the comparison between our
observational data and the new PHOENIX models are summarised in
Table\,\ref{tab:results}. We visualise our results in comparison to
literature data for effective temperature in the top panel of
Fig.\,\ref{fig:parcompare}, $\log{g}$ in the middle panel, and
metallicity in the bottom panel of Fig.\,\ref{fig:parcompare}.

\subsection{Effective temperature and surface gravity}

Our results for temperature and surface gravity are consistent with most
of the literature data within 1\,$\sigma$ uncertainties (see top panel
in Fig.\,\ref{fig:parcompare}).  For GJ~551, the temperature
  from \cite{Neves2013} is significantly cooler than our result and
  the one from \cite{Segransan2003}. This discrepancy may be caused by
  the TiO bands starting to saturate at cooler temperatures, as
  mentioned in Sect. 3.3. This consistency is crucial for our test
of the accuracy of the new PHOENIX ACES model generation. It shows
that with the new model set we are able to obtain physically
meaningful parameters of very cool stars. Furthermore, the fact that
the synthetic spectra match the TiO bands as well as atomic lines
with one consistent parameter set (see Fig.\,\ref{fig:stdM_comp})
indicates that the microphysics used for the new PHOENIX atmosphere
grid has improved significantly with respect to older model
generations \citep{Reiners2005}. For GJ~411 and GJ~699 our 
results perfectly agree with those of \citet{Mann2015} within 
10~K. 

\subsection{Metallicity}

For the three benchmark stars used above, GJ~411, GJ~699, and GJ~551, 
direct radius measurements
provide independent information on temperature and surface
gravity. The third crucial atmospheric parameter, metallicity, is not
independently constrained, but can only be determined from
spectroscopic analysis. Literature values given in
Table\,\ref{tab:results} are also based on spectroscopic models, which
are often calibrated using components of binary stars, as discussed in
the introduction. For the three stars GJ~411, GJ~699, and GJ~551, our
metallicities are not fully consistent with the literature
values of \cite{Neves2013}; on average, our results are $\Delta$[Fe/H]~=~0.22\,dex higher.
As an independent test of our metallicity results, we consider
the M1.5 dwarf GJ~667C, which is part of a triple system and often
used to anchor M-dwarf metallicity scales. For the brighter
components, [Fe/H] is very accurately determined with a mean
literature value of [Fe/H]~=~-0.55 (Sect.\,\ref{sect:objects}).
This low-metallicity star should be a good indicator for whether our models can determine metallicities in M dwarfs with
  sub-solar metal abundances. Our result is [Fe/H]~=~$-0.56\pm$0.11,
  which agrees very well with the metal abundance of components A and
  B and also with the result from \cite{AngladaEscude2013} (who used the same method). The value reported by
\cite{Neves2014} is [Fe/H]~=~$-0.5\pm$0.09, which is also consistent with the
system's higher mass components.

We conclude from this first test that the new PHOENIX ACES models
provide reliable results in $\log{g}$ and $T_{\rm eff}$. For
metallicity, the situation is more difficult. Our models indicate
higher metallicities than other methods in the three stars for which
temperature and gravity are constrained from knowledge about the
stellar radii. In the fourth object, which has information on metal
abundance from its higher mass companions, our methods provide
results that are consistent with other methods. Further investigation including
comparison of metallicities in other binary systems will be carried
out in a subsequent paper when a larger M-dwarf sample will be
investigated.

\section{Summary and discussion}

We have determined atmospheric parameters from high-resolution optical
spectroscopy in a few benchmark M dwarfs with the new PHOENIX ACES
models. The models include an improved description of physical
processes in low-temperature atmospheres and are expected to provide a
better fit to spectroscopic observations and accurate atmospheric
parameters. We identified useful spectroscopic indicators with which
the three parameters can be constrained within small
statistical uncertainties. Uncertainties due to degeneracies
between spectra with different parameters are $\sigma_{T_{\rm eff}} =
35$\,K, $\sigma_{\log{g}} = 0.14$, and $\sigma_{[Fe/H]} = 0.11$. 

The main purpose of the paper was a comparison of the model spectra to
stars with independent information on atmospheric
parameters. Temperature, surface gravity, and metallicity were
determined for four stars. For three of the four, we compared our
model results to temperature and gravity known from interferometric
radius measurements, luminosity, and parallax. In all six parameters,
the agreement is excellent (within 1\,$\sigma$). Metallicity
measurements are also available for the three stars, but they
are inconsistent
with our results. On average, our model fits are 0.31\,dex
higher than literature values, which is similar to the offset found by
\citet{JohnsonApps2009}. A comparison with results in the NIR
\citep{RojasAyala2012} showed an average deviation of 0.33 dex for
GJ411 and GJ699. The same difference was found when comparing 
our results for these two stars to \citet{Mann2015}. At this point, 
we cannot clearly identify the
reason for this discrepancy. One possible explanation is that our
improved models provide a better description of the cool atmospheres
and therefore more accurate metallicities than other methods. 
However, we cannot exclude the possibility that our metallicities systematically overestimate the metal abundances.

\cite{Rajpurohit2014} determined stellar parameters of M dwarfs 
using BT-Settl models. They reported good fitting of the TiO-bands, but 
rather poor fitting of the K- and Na-line pairs for later-M dwarfs, with 
the observed lines being broader and shallower than the models. We 
observe a similar behavior (see spectra for GJ~551 and GJ~699 in 
Fig. \ref{fig:stdM_comp}), although in our case the models show broader 
lines. However, this might be an indication that current models still have 
problems to reproduce lines in the lower temperature and maybe lower 
metallicity range.

\citet{Pavlenko2015} investigated the binary system G~224-58~AB, 
consisting of a cool M extreme subdwarf and a brighter K companion. 
They determined abundances of different elements for the bright component 
using synthetic spectra calculated with the WITA6 program \citep{Pavlenko1997} 
and checking the agreement to the spectrum with models with different 
parameters. Abundances for the cooler M companion were determined 
independently by using NextGen and BT-Settl models \citep{Hauschildt1999}. 
No comparison with literature values was made, therefore the validity of the abundances 
cannot be verified. However, they proved the binarity of the system by showing 
that both components have the same metallicities derived by different methods.
This shows that even previous-generation models were capable of reproducing 
metallicities using specific lines, which indicates that the new-generation models 
might work even better using whole molecular bands.

To address our offset in metallicity, we investigated a fourth
star that is a member of a multiple system with known
metallicities. In this case, our result agrees with the literature
values reported for the more massive components, in which
determinations of metallicity are better established. While this
comparison supports the interpretation that the new models accurately
describe cool atmospheres, it does not solve the puzzle of why our
metallicities disagree with other M-dwarf metallicity scales because
in this particular star other determinations (are designed to) agree
as well.

High-resolution spectra of low-mass stars can potentially be used to
determine atmospheric parameters and even individual element
abundances to high accuracy. We have shown that the new PHOENIX grid
provides a good set of models to start such a comparison. We plan to
use our method to derive atmospheric parameters from many M-dwarf
spectra, and this analysis should also include a comparison to
other metallicity scales. If the real metallicities of M dwarfs are
indeed several tenths of a dex higher than currently assumed, this
would have serious ramifications for our understanding of planet
formation and the local stellar population. Therefore, a consistent
understanding of \emph{all} spectral features in cool atmospheres is
mandatory.



\begin{acknowledgements}
  We thank Barbara Rojas-Ayala and Andreas Schweitzer for
    fruitful discussions about low-mass star atmospheres and
    metallicity scales. VMP acknowledges funding from the GrK-1351
    \emph{Extrasolar Planets and their Host Stars}. SWvB acknowledges
  DFG funding through the Sonderforschungsbereich 963
  \emph{Astrophysical Flow Instabilities and Turbulence}, and the
  GrK-1351 \emph{Extrasolar Planets and their Host Stars}.  AR
  acknowledges funding from the DFG as a \emph{Heisenberg}-Professor
  under RE-1664/9-2 and support by the European Research Council under
  the FP7 Starting Grant agreement number 279347.
\end{acknowledgements}

\bibliographystyle{aa}  
\bibliography{wende.bib}

\end{document}